# Simple structured illumination microscope setup with high acquisition speed by using a spatial light modulator

**Ronny Förster,**[1,2] **Hui-Wen Lu-Walther,**[1] **Aurélie Jost,**[1,2] **Martin Kielhorn,**[1,3] **Kai Wicker,**[1,2,*] **and Rainer Heintzmann**[1,2,3,**]

[1] *Leibniz Institute of Photonic Technology, Jena, Germany*
[2] *Institute of Physical Chemistry, Center of Photonics, Friedrich-Schiller-University Jena, Germany*
[3] *Randall Division of Cell and Molecular Biophysics, King's College London, London, UK*
*(\*currently at Carl Zeiss AG, Corporate Research and Technology, Jena, Germany)*

*\*\* heintzmann@gmail.com*

**Abstract:** We describe a two-beam interference structured illumination fluorescence microscope. The novelty of the presented system lies in its simplicity. A programmable electro-optical spatial light modulator in an intermediate image plane enables precise and rapid control of the excitation pattern in the specimen. The contrast of the projected light pattern is strongly influenced by the polarization state of the light entering the high NA objective. To achieve high contrast, we use a segmented polarizer. Furthermore, a mask with six holes blocks unwanted components in the spatial frequency spectrum of the illumination grating. Both these passive components serve their purpose in a simpler and almost as efficient way as active components. We demonstrate a lateral resolution of 114.2 ± 9.5 nm at a frame rate of 7.6 fps per reconstructed 2D slice.

## 1. Introduction

Fluorescence microscopy is a well-established method in life science. Unfortunately, its resolution is fundamentally limited by diffraction and details that are smaller than the diffraction limit are unresolved. The resolution of a classical microscope is defined by $\lambda/2n \times \sin\theta$ in the lateral plane and $\lambda/n \times (1-\cos\theta)$ in the axial direction [1]. There are several methods to improve the resolution of wide-field microscopy [2–7].

One of them is structured illumination microscopy (SIM). Here, a periodically modulated illumination pattern is projected into the sample. The whole region of interest is acquired in one image. Thus, SIM has the capability of super resolution with high frame rates even in a large region of interest (ROI), because it is a wide-field technique.

In SIM, objective information with a high spatial frequency is down-modulated by the frequency of the illumination grating in the Fourier domain. Thus, frequencies higher than the Abbe limit can be shifted into the pass-band of the optical transfer function (OTF) and are transferred by the microscope. The resolution enhancement is limited to a factor of two, in the case where the illumination grating itself is generated by the objective [8, 9]. Several raw images for different pattern positions as well as image processing are necessary to undo the frequency modulation and to compute the final high-resolution image out of the raw SIM images [10]. This procedure has to be repeated in several directions because the down modulation is along one direction only. If a living cell moves between individual raw images, the reconstruction will exhibit artifacts. The shorter the acquisition time, the fewer motion artifacts will be in the final image.

The reconstruction algorithm needs precise knowledge about the shape of the illumination pattern. In our setup a sinusoidally modulated pattern is used. This is achieved by the interference of two plane waves (two-beam interference) originating from two focused spots in the back focal plane (BFP) of the objective. These are the plus and minus first diffraction orders of a grating positioned in an intermediate image plane. As a result, the fundamental sinusoidal component of the diffraction grating is projected into the sample plane.

Lateral and axial resolution can be further enhanced by illuminating with different patterns. If the zero diffraction order is also used (three-beam illumination), the illumination pattern is modulated additionally along the axial direction. With this illumination, axial and lateral resolution can be improved by a factor of two compared to wide-field. Furthermore, optical sectioning is achieved by filling the missing cone.

Further improvements have been made to the SIM technique by the invention of non-linear SIM (NL-SIM) [12]. In this case, saturation processes lead to an effective sub-diffraction illumination grating, thus enabling only noise limited resolution. Rego *et al.* have demonstrated a resolution of 50 nm for fixed nuclear pore complexes[13]. However, the number of raw images necessary for the reconstruction is increased nearly by one order of magnitude. Therefore, acquisition speed and frame rate are substantially decreased.

Current Fast-SIM setups achieve an accurate illumination pattern and precise phase stepping, by controlling the illumination beam is using a number of active components [11]. These include a spatial light modulator (SLM) for pattern stepping, an active pupil aperture for blocking unwanted diffraction orders and an active polarization rotator for ensuring high pattern contrast. All of these need a precise characterization, before the challenging alignment can be done. Some devices are not commercially available but are custom-built. Additionally, synchronization is necessary. Based on the advanced setup by Fiolka *et al*. [11], we replaced complex active devices by unsophisticated passive components. Therefore, our setup needs less synchronization and is simplified.

The speed of our system makes it an ideal candidate for NL-SIM. We discuss if our newly introduced components can theoretically fulfill the more stringent demands of NL-SIM.

**Existing systems**

Although SIM is a wide-field technique, it is slower than conventional microscopy because several raw images for different illumination patterns have to be acquired. Shifting and rotating the projected grating by moving the diffraction grating in the intermediate image plane itself is time-consuming. In commercial systems, like the Elyra S.1 (Zeiss, Germany), it takes approximately 1.4 s to acquire a single plane 2D high resolution image (512 × 512 pixel; three-beam illumination) [14]. Thus, it is prone to motion artifacts and it is impossible to record biological processes which happen in less than a second.

Kner *et al*. published a method to reduce the acquisition time by producing the gratings with a spatial light modulator (SLM). The SLM can switch faster and more precisely between two different illumination grating positions than a physical grating, because the displayed pixel values simply has to be changed [15]. However, replacing the diffraction grating with the SLM is complex. Some additional elements have to be inserted to the setup to remove the drawbacks [16].

The grating displayed on the SLM is pixelated and binary. Therefore, its Fourier transform exhibits more than the two desired first diffraction orders. To have a sinusoidal illumination pattern in the sample, the additionally appearing 'unwanted orders' have to be blocked in a pupil plane. The approach from Fiolka *et al*. is a "rotating-slit". It is a narrow mechanical slit, centered on the zero order. The slit is rotated to the direction of the first orders so that they can pass. The filtering always works perfectly because the patterns displayed on the SLM are designed in such a way that no unwanted order can lie on the slit, as theory shows. The drawbacks of the slit are that it is not commercially available, needs to be synchronized and consumes 1.5 ms to rotate to the next direction [11].

A second issue is that only the azimuthally polarized components of the two beams interfere and form a grating. The rest forms an offset and is lost for the application of SIM. The interference of azimuthally polarized orders never has an offset and has therefore a perfect contrast and the best signal-to-noise ratio (SNR) for all grating directions. Unfortunately, the diffracted orders from the SLM all have the same linear polarization [16]. Fiolka *et al*. used an active "polarization-rotator" (SWIFT, Meadowlark, USA) to rotate the polarization in each order into the specific direction. The device needs a very precise adjustment, synchronization and ~1 ms to get ready for the next polarization direction [11].

Polarization-rotator and rotating-slit perform very well in practice. Both devices correct the drawbacks of the SLM and lead to a sinusoidally modulated illumination pattern of high contrast [11].

**2. Materials and Methods**

We have built a simple Fast-SIM setup, as depicted in Fig. 1. The design is a simplified version of the setup of Fiolka *et al*. [11].

The excitation laser (442 nm/70 mW) is controlled by an acousto-optical tunable filter (AOTFnC-VIS, AA Opto-Electronic, France). A beam collimator consisting of a pinhole (Ø= 25μm, Thorlabs, USA) and two lenses (f1 = 35 mm and f2 = 100 mm, Thorlabs, USA) achieves mode cleaning and beam expansion.

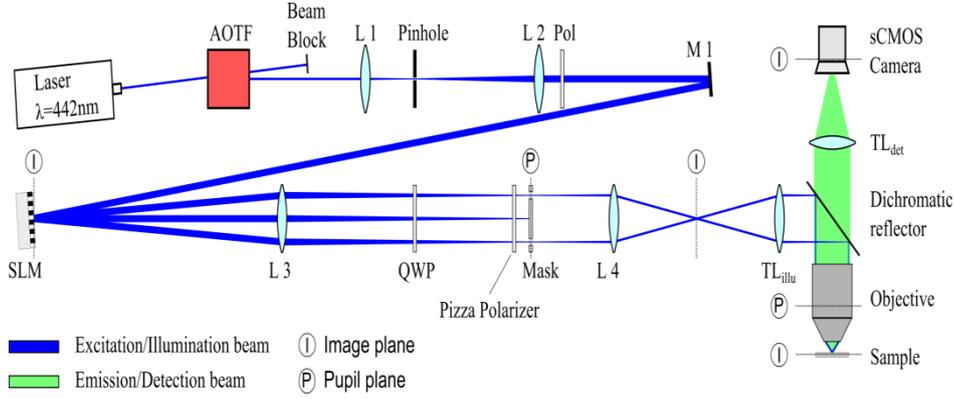

Fig. 1. The AOTF switches the illumination beam on and off. Mode cleaning and beam expansion is performed by the lenses L1 & L2 and a pinhole. The SLM is illuminated and the light is diffracted into different orders. L3, L4, the illumination tube lens and the objective project the grating from the SLM in the sample plane. A "pizza" polarizer and a mask accomplish a sinusoidal illumination grating of high contrast in the sample. Emitted light goes through the dichromatic reflector and is imaged with the detection tube lens to the sCMOS camera.

The SLM (SXGA-3DM, Forth Dimension Displays, UK) is placed in an image plane. It displays a grating which diffracts the light [16]. Appendix A explains the grating properties and how the precise patterns designed, based on the ideas from Shao *et al*. [17]. The SLM used has the common low diffraction efficiency even for the optimum green light (532 nm). Using blue light instead further reduces the amount of light in the desired first orders to 3 %.

A polarizer in the illumination beam (22 CA 25, Comar, UK) ensures that the diffracted orders are also linearly polarized [16]. This is necessary for the subsequent polarization control. The beam illuminates the SLM under an angle of incidence of 2°. This separates the diffracted beams from the illumination beam. A beam splitter is not required [15]. In the setup, the beam diameter is roughly one third of the shorter side of the SLM. This ensures that the beam does not illuminate the scattering mounting of the SLM. On the other hand, the beam is big enough to illuminate a large field of view in the sample.

The spatial frequency domain of the displayed grating is projected into the BFP of lens 3 (f3 = 750 mm, Thorlabs, USA). Here we use a mask to block unwanted orders (see Appendix A). It is a fixed cardboard with six little holes at the positions of the desired first diffraction orders [15]. We verified in a simulation that no unwanted orders can pass the mask through one of the holes. Thus, our mask filters as efficiently as the rotating slit. But unlike the rotating slit can only accommodate the diffraction orders for a single fixed wavelength. However, using the mask saves effort and time.

Two passive optical components guarantee azimuthally polarized light. The first one is a quarter wave plate (QWP), which realizes circular polarization in all first diffraction orders. The second element is a patterned polarizer (Codixx, Germany). This polarizing plate is composed of twelve identical angular segments (six would be enough for our application). The transmission axis of each individual section is orthogonal to its chord. The result is the nearly azimuthal polarizer of Fig. 2. Because of its shape, we termed it "pizza polarizer". The pizza polarizer is located close to a pupil plane where the diffraction orders are focused and traverse the right segment. To avoid scattering of unwanted orders on the glue which connects two segments, the pizza polarizer should be located behind the mask. Due to a lack of space we had to place them the other way round. However, we saw no negative effect of this order in the experiment. Half of the light is lost using this combination of QWP pizza polarizer. However, it is easy to adjust, does not require any synchronization, control or loading time.

Lens 4 (f4 = 150 mm, Thorlabs, USA) and the illumination tube lens ($f_{TL-Illu}$ = 125 mm, Thorlabs, USA), image the pupil plane into the BFP of the objective. The objective (Plan-Apochromat ×63/1.4 oil DIC, Zeiss, Germany) generates the two-beam interference in the sample plane. The grating constant in the sample is 193 nm. The field of view in -sample space is 19.5 × 13.8 µm². The emitted fluorescent light is imaged by the objective and the detection tube lens ($f_{TL-det}$ = 164.5 mm, Zeiss, Germany) onto a modern camera with sCMOS technology and fast readout (Orca Flash 2.8, Hamamatsu, Japan). We run the camera in the 'global exposure level trigger' mode, which operates with a time-saving permanently opened shutter. The AOTF prevents the camera from being illuminated during readout.

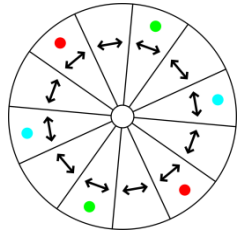

Fig. 2. Twelve circular segments of linear polarizers form one azimuthally polarizer – termed pizza polarizer. The position of the two first orders is shown for each grating direction in different colors.

Camera, AOTF and SLM are synchronized by a microcontroller (Arduino Uno, Arduino, Italy). Thus, for each raw image, the same sample area is illuminated with a different grating. The camera integrates when the SLM displays a grating. In order to prevent damage of the liquid crystals (by ionization), the SLM has to display a grating and its inverse successively before loading the next grating. In case of two-beam illumination, the grating and its inverse lead to the same projected pattern in the sample. Therefore, each raw image is illuminated by both. The AOTF switches off the illumination beam while the SLM is loading a grating image which takes ~0.434 ms

Due to the losses in the optical components and diffraction effects polarization design, less than 1 % of the laser output intensity reaches the sample. Additionally, the laser wavelength is not at the excitation maxima of the used fluorophores. Thus, an exposure time of 8 ms per raw image was necessary to have an acceptable SNR in the image. The SLM can display a grating only up to 2 ms before harmful effects are generated in the SLM. Therefore, the SLM displays each grating four times for 2 ms (two positive-negative image pairs). Furthermore, the SLM displays a non-illuminated grating during the readout of the camera. The SLM runs continuously, as we had the impression that the synchronization was disturbed if the SLM stopped displaying gratings during read out.

**3. Results**

We took a picture of the pupil plane with a simple web camera Fig. 3. The left side shows this plane without any filtering. The unwanted orders are clearly visible. The right image shows the same plane after insertion of the mask. No unwanted orders pass the mask through one of its six holes making it suitable for all three grating directions. The simple idea of a mask works quite well.

We illuminated with 110 nm beads sample via the grating. Fig. 4 shows the image of one single bead. While shifting the grating laterally, we observed that the emission is bright when a bead is on a maximum intensity (left) and very dim when it is at the minimum (right). This demonstrates the high contrast of the illumination grating. Note that the bead is not completely extinguished on the right because it is wider than the dark fringe itself, and therefore still slightly excited.

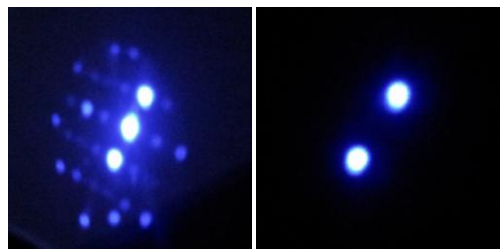

Fig. 3. Left: Unwanted orders are present in the pupil plane of the SLM. Right: The used mask blocks them. (The perspective leads to distortion.)

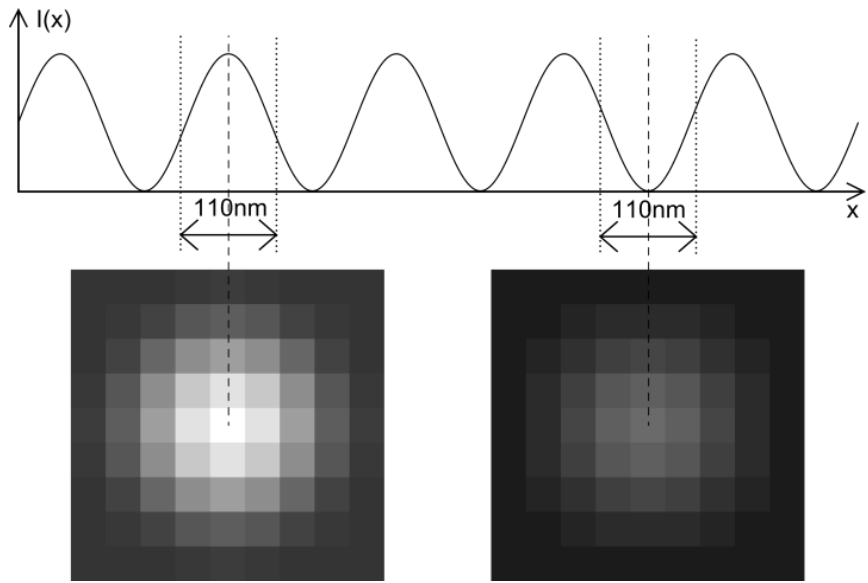

Fig. 4. 110 nm bead with different brightness resulting from the high illumination grating contrast (left: maximum, right: minimum). The gray values are non-linear (gamma=0.4).

To demonstrate and characterize the performance of our Fast-SIM system in practice, we imaged a sample consisting of 110 nm fluorescent beads. Fig. 5 shows a comparison of conventional wide-field microscopy with SIM, which has a clearly improved lateral resolution (for a fair comparison, linear image deconvolution has been applied to both images. [18]). Fig. 6 shows a profile along the white line in Fig. 5. SIM clearly resolves the three beads on the right (green), while they are nearly unresolved in wide-field (blue dotted). On top of that, edges are steeper in SIM than in wide-field.

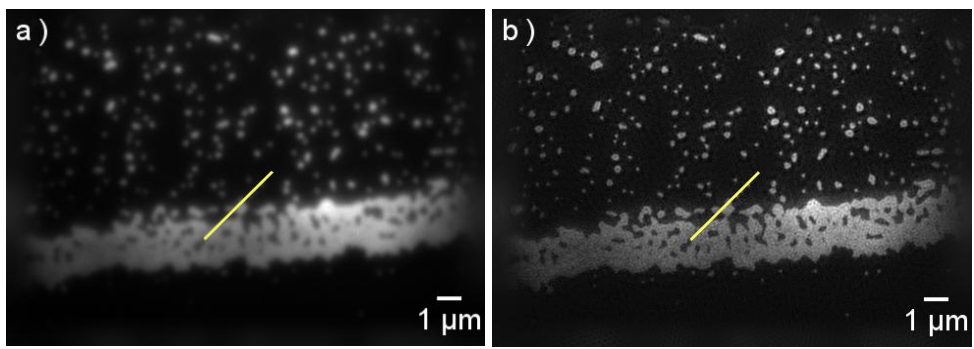

Fig. 5. Comparison of wide-field (a) and SIM (b) with a 110 nm bead sample

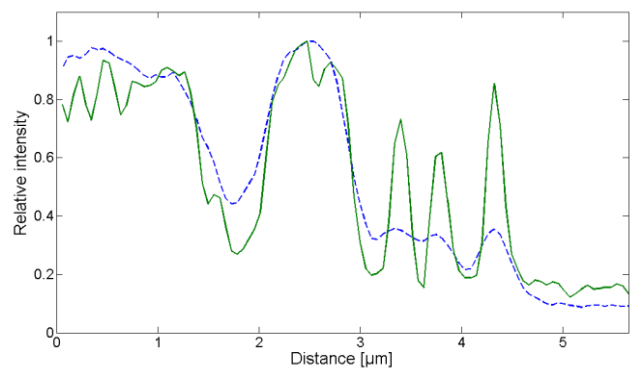

Fig. 6. Wide-field (blue dotted) and SIM (green) profile along the white dotted line from Fig. 5 beginning at the left bottom.

50 beads that appear bright and well-isolated in wide-field and SIM were selected in both wide-field and SIM images to determine the full width at half maximum (FWHM) of the effective PSF (Fig. 7). This average bead size in wide-field image has an FWHM of 224.8 ± 21.2 nm. In SIM, the FWHM is reduced to 114.2 ± 9.5 nm. The enhancement could be improved further, by bringing the grating constant (currently 193 nm) closer to the diffraction limit ( $\lambda_{ex}$ / (2×NA) = 442 / (2×1.4) = 158 nm).

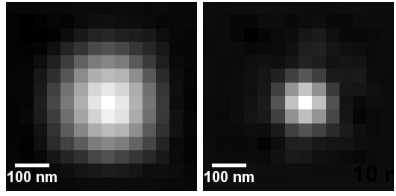

Fig. 7. Average point spread function of wide-field (left) and SIM (right) from Fig. 5

The acquisition time for one raw image is 14.6 ms, i.e. 68.5 frames per second (fps). The final high-resolution image rate is the ninth part - 7.6 fps. The setup is ten times faster than the commercial system Elyra S.1, which provides 0.7 fps.

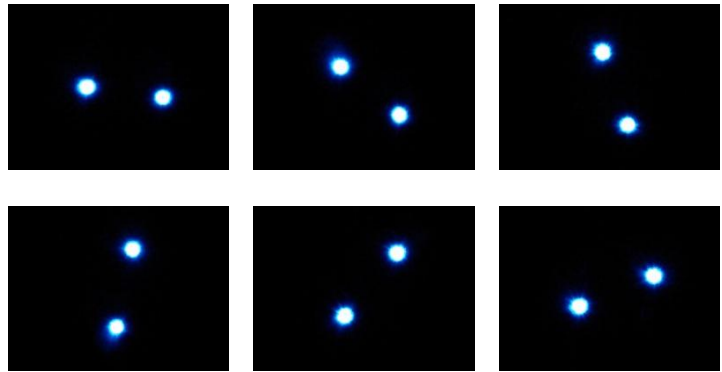

Fig. 8. Six different grating directions filtered by a mask in a pupil plane. All unwanted orders are blocked.

## 4. Discussion

We have built a Fast-SIM system which uses passive and unsophisticated components for selecting the illumination orders and controlling their polarization. The use of a passive mask for order selection requires the calculation of unique grating patterns, which guarantee the separation of wanted from unwanted orders between all orientations. Since we have found a triplet of matching gratings, our mask perfectly filters the illumination beam and made the rotating slit method superfluous. However, a rotating slit has potential advantages when working with multiple wavelengths as otherwise the holes in our mask will have to be elongated or repeated for each wavelength leading to potential unwanted oder crosstalk. The "pizza" polarizer guarantees in a very simple way for an illumination pattern of high contrast for all directions. Exposure time or laser power has to be increased because of overall low transmittance. The polarization rotator has a higher transmittance. The company asserts more than 90 % [19]. Thus, the illumination efficiency is higher if the polarization rotator is used. Besides, it has a better performance in three-beam illumination, because it rotates the zero orders similar to the first orders. The clear center of the pizza polarizer leaves the center circularly polarized. Thus, three-beam illumination is possible, however with a reduced grating contrast and SNR. The greatest advantage using a mask and a pizza polarizer is simplifying the system setup and synchronization.

We demonstrate that our setup is able to achieve superresolution by imaging beads with a FWHM of 114.2 ± 9.5 nm at 7.6 fps. The exposure time of 8 ms per raw image is comparatively long. We believe that the exposure time can be reduced to 1 ms by using better matching fluorophores and a laser with higher power. In that case the raw data frame rate can be doubled to 132 fps. This corresponds to a final image rate of 14.7 fps.

## 5. Outlook

The setup can be expanded to a nonlinear SIM system. Rego *et al*. used nine grating directions to achieve an isotropic resolution of 50 nm [13]. We think that using six directions will be sufficient for this purpose. We therefore extended our grating search algorithm by three directions. Additionally, we updated the grating design for even smaller and more precise gratings [11]. We found six gratings which can be filtered by a mask (see Fig. 8) and have a grating period of roughly 39 µm on the SLM (reduction of 58 %). This will enlarge the ROI over five times and the system will be shortened by the higher diffraction angle on the SLM. The deviations from the desired 30° angle (180° / 6 directions) are less than 1°. The algorithm and the new gratings are attached in Appendix B. NL-SIM needs a high illumination contrast [13]. This can be provided by our pizza polarizer with 12 segments.

The use of passive components for order selection and polarization control significantly simplifies the setup. In order to facilitate the construction of such a home-built system for other labs, we provide detailed instructions on how to synchronize the remaining components and generate patterns suitable for passive filtering in the supplementary materials.

# Appendix

## A. Grating design (after Shao *et al.*)

The gratings used in the experiments are designed and characterized based on the ideas from Shao *et al.* [17]. Fig. 9 shows an example grating. White pixels are in the on state and black pixels are off. The 2D periodic pixel pattern forms a grating.

The theory of crystal structures with lattice vectors is suitable to describe the patterns [20]. The lattice points are marked in red. The lattice vectors $\vec{a}$ connect two lattice points. Two different lattice vectors describe the two-dimensional grating pattern P completely:

$$P = \{\vec{a_H}; \vec{a_\theta}\} \quad \text{with:} \quad \vec{a_H} = \begin{pmatrix} h \\ 0 \end{pmatrix} \quad \text{and} \quad \vec{a_\theta} = \begin{pmatrix} \theta_x \\ \theta_y \end{pmatrix} \tag{1}$$

The lattice vector $\vec{a_H}$ is always horizontal and defines the horizontal grating period. The grating direction is only defined by $\vec{a_\theta}$ (see blue dotted line in Fig. 9). The range of search for a useful grating is given by:

$$h \in \{1, 2, \ldots, h_{max}\}$$

$$\theta_x \in \{0, 1, \ldots, \theta_{x_{max}}\}$$

$$\theta_y \in \{-\theta_{y_{max}}, -\theta_{y_{max}} + 1, \ldots, \theta_{y_{max}} - 1, \theta_{y_{max}}\} \tag{2}$$

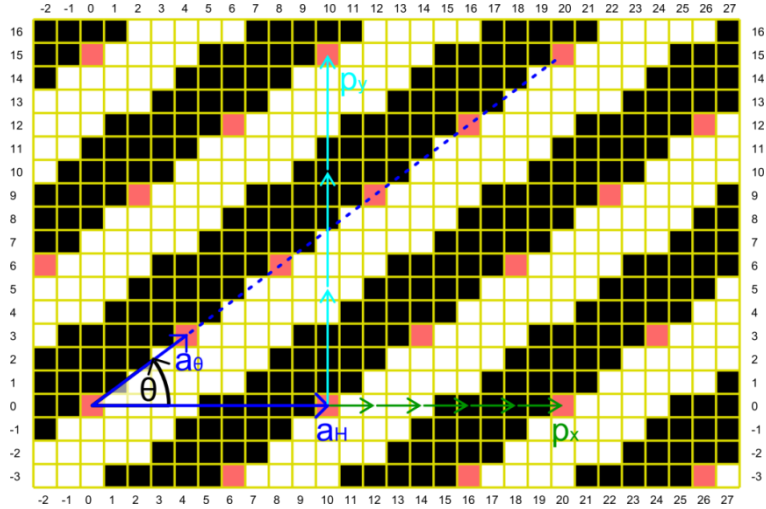

Fig. 9. Grating on a SLM. White and red pixels are on, black pixels are off. The red pixels are lattice points, which are connected by the lattice vectors $\vec{a_\theta}$ and $\vec{a_H}$. The grating direction $\theta$ is only defined by $\vec{a_\theta}$. The grating constant has to be calculated with both lattice vectors. Five equidistant phase steps (three-beam illumination) are provided by a horizontal shift (green). Three such phase steps (two-beam illumination) are done vertically (cyan).

The characteristic grating parameters, orientation $\theta$ and grating period $g$, can be calculated with:

$$\theta(P) = \theta(\vec{a_\theta}) = \begin{cases} \dfrac{\pi}{2} & \text{for:} \quad \theta_x = 0 \\ \arctan\left(\dfrac{\theta_y}{\theta_x}\right) \in [0; \pi] & \text{for:} \quad \theta_x \neq 0 \end{cases} \tag{3}$$

$$g(P) = g(\vec{a_H}, \vec{a_\theta}) = h \cdot \sin(\theta) \tag{4}$$

Three gratings P form a triplet $T = \{P_1; P_2; P_3\}$. A brute-force algorithm searches systematically for three matching gratings. Therefore, the triplet has to pass three tests before being accepted.

1. The angles α(T) between the gratings must be close to 60°. We accept a triplet if the angular deviation is smaller than 5°:

$$\alpha(T) = \{|\theta(P_1) - \theta(P_2)|; |\theta(P_1) - \theta(P_3)|; |\theta(P_2) - \theta(P_3)|\}$$
$$\Delta\alpha(T) = \max\{|\theta(T) - 60°|\} < 5° \quad (5)$$

2. All three grating constants should be similar. To check this, we define the average grating constant of the triplet g(T). The mismatch Δg(T) between a single grating constant and this average value has to be smaller than 2%:

$$g(T) = \{g(P_1); g(P_2); g(P_3)\} \quad (6)$$

$$\overline{g}(T) = \frac{\max(g(T)) + \min(g(T))}{2} \quad (7)$$

$$\Delta g(T) = \frac{\max\{|\overline{g}(T) - g(T)|\}}{\overline{g}(T)} < 2\% \quad (8)$$

3. The mask, as a Fourier filter in the experiment, can only operate if no unwanted order is at the position of any desired first diffraction order. Therefore, we calculate the Fourier transform of all three gratings and assign each of them to one channel of a RGB-image. A gamma correction of 0.3 is applied to enhance the visibility of the weaker unwanted orders. Six yellow circles around the first diffraction orders represent the holes in the mask. We checked manually that no unwanted orders are in the circle – see Fig. 10. The white spot in the middle is the zero order.

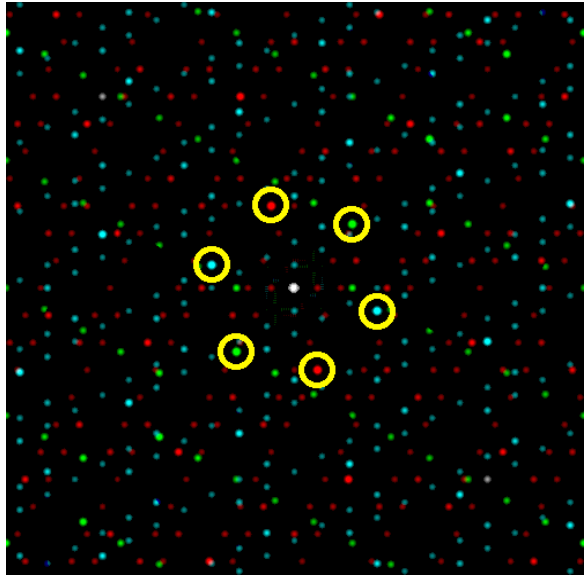

Fig. 10. All three Fourier tranforms of the gratings in one RGB image. The yellow circles represent the holes in the mask. The triplet is useless if unwanted orders are in the circle. Gamma correction of 0.30 is applied for a better visibility.

An additional step can test the possibility of homogeneous phase steps. In this case, the intensity patterns of all grating steps add up to a homogeneous distribution. Thus the illumination beam does not bleach any grating into the sample in time. We found out that homogeneous phase steps occur, when h or $\theta_y$ are multiples of the number of phase steps. Unfortunately, this advantage significantly reduces the number of matching gratings a lot. We do not have homogeneous phase step because it is not necessary for the reconstruction algorithm and our laser is not powerful enough to bleach a grating into the sample.

In our experiment we used the following three grating parameters:

|  | Grating 1 | Grating 2 | Grating 3 |
|---|---|---|---|
| h [Px] | 7 | 10 | 25 |
| $\theta_x$ [Px] | 7 | 10 | 25 |
| $\theta_y$ [Px] | 25 | -9 | 7 |
| $\bar{g}(T)$ [Px] | $6.72 \pm 0.03 \rightarrow \Delta g(T) = 0.4\%$ | | |
| $\bar{g}(T)$ [Px] | $91.5 \pm 0.4$ | | |
| $\theta$ [°] | 74.4 | 138.0 | 15.6 |
| $\alpha(T)$ [°] | 63.6 | 57.6 | 58.8 |

## B. Future grating design (after Rego *et al.*)

The new gratings are designed and characterized based on the ideas from Rego et al. and intended for NL-SIM [11]. Fig. 11 shows an example grating. White and red pixels are in the on state and black pixels are off.

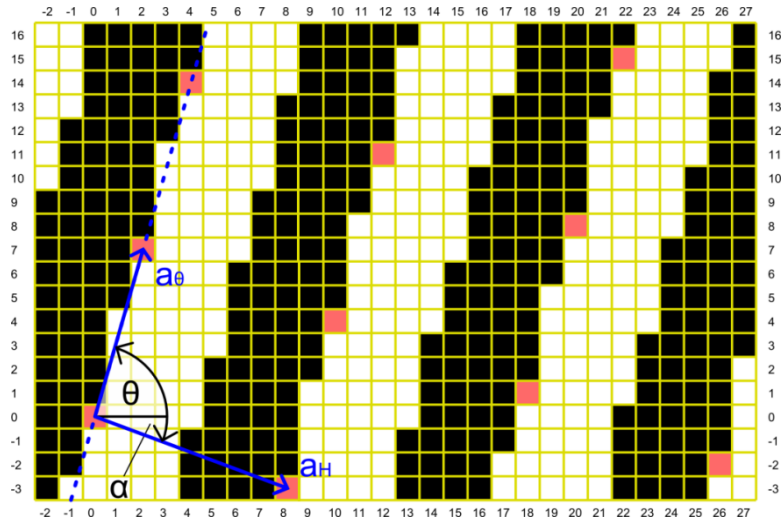

Fig. 11. Grating on the SLM based on Rego's approach. White and red pixels are on, black pixels are off. The red pixels are lattice points, which are connected by the lattice vectors $\vec{a_\theta}$ and $\vec{a_H}$. The grating direction is still only defined by $\vec{a_\theta}$. The second lattice vector $\vec{a_H}$ does not have to be horizontal anymore.

Two lattice vectors completely describe the two-dimensional grating pattern P:

$$P = \{\vec{a_H}; \vec{a_\theta}\} \quad \text{with:} \quad \vec{a_H} = \begin{pmatrix} h_x \\ h_y \end{pmatrix} \quad \text{and} \quad \vec{a_\theta} = \begin{pmatrix} \theta_x \\ \theta_y \end{pmatrix} \quad (9)$$

There exist more gratings to choose from, because $\vec{a_H}$ is not restricted anymore to be horizontal. Grating orientation θ and period g can be calculated with:

$$\theta(P) = \theta(\vec{a_\theta}) = \begin{cases} \frac{\pi}{2} & \text{for: } \theta_x = 0 \\ \arctan\left(\frac{\theta_y}{\theta_x}\right) \in [0; \pi] & \text{for: } \theta_x \neq 0 \end{cases} \quad (10)$$

$$g(P) = \sqrt{h_x^2 + h_y^2} * |\sin(\theta(\vec{a_\theta}) - \theta(\vec{a_H}))| = \frac{h_x\theta_y - h_y\theta_x}{\sqrt{h_x^2+h_y^2}} \quad (11)$$

The algorithm is extended to six directions. It is still a brute-force algorithm, which searches systematically for the matching gratings. Each sextet $S = \{P_1; P_2; P_3; P_4; P_5; P_6\}$ has to pass four tests before being accepted. The angles α(S) between the gratings have to be close to 30°. We accept a sextet if the angular deviation is always smaller than 1°. All six grating constants should be similar. We define the average grating constant of the sextet g(S) again as the mid-range value. The mismatch Δg(S) between a single grating constant and this average value has to be smaller than 1%. The algorithm automatically checks if a mask can block all unwanted orders. A square of several pixels is defined around each first order. The intensity of all unwanted orders in this square is summed and compared to that of the first order. The sextet is rejected if the intensity of the first order is not at least 100 times that of the unwanted orders.

Homogeneous phase steps are favorable in non-linear SIM because the saturation processes might lead to bleaching artifacts. Seven phase steps enable the reconstruction of the first and second higher-order harmonic to double the lateral resolution of SIM [13]. We check if either a horizontal or vertical phase step is possible (*mod* – modulo and *lcm* – least common multiple):

Vertical phase step test ($\theta_x \neq 0$ required):

If: $h_x = 0$: $\qquad h_y \bmod 7 = 0$

Else: $\left(lcm(|h_x|, |\theta_x|) * \frac{\theta_y}{\theta_x} - \frac{h_y}{h_x}\right) \bmod 7 = 0$ \quad (12)

Horizontal phase step test ($\theta_y \neq 0$ required):
If: $h_y = 0$: $\qquad h_x \bmod 7 = 0$

Else: $\left(lcm(|h_y|, |\theta_y|) * \frac{\theta_x}{\theta_y} - \frac{h_x}{h_y}\right) \bmod 7 = 0$ \quad (13)

The sextet we used for our first experiments is:

|  | Grating 1 | Grating 2 | Grating 3 | Grating 4 | Grating 5 | Grating 6 |
|---|---|---|---|---|---|---|
| $\theta_x$ [Px] | 1 | 13 | 24 | 27 | 21 | 13 |
| $\theta_y$ [Px] | -27 | -21 | -13 | 1 | 13 | 24 |
| $h_x$ [Px] | 2 | 12 | 7 | 4 | 14 | 6 |
| $h_y$ [Px] | 23 | -14 | -7 | 3 | 12 | 17 |
| g(S) [Px] | 2.850 | 2.834 | 2.821 | 2.850 | 2.834 | 2.821 |
| $\overline{g}(S)$ [Px] | $2.835 \pm 0.015 \rightarrow \Delta g(T) = 0.6\%$ | | | | | |
| $\overline{g}(S)$ [µm] | $38.61 \pm 0.21$ | | | | | |
| θ [°] | 2.12 | 31.76 | 61.56 | 92.12 | 121.76 | 151.56 |
| $\Delta\alpha(S)$ [°] | **29.64** | **29.80** | **30.56** | **29.64** | **29.80** | **30.56** |

### C. SLM software

Our SLM (SXGA-3DM) is controlled via the Software MetroCon 2.0 from ForthDD. First, the user has to upload his gratings. Afterwards, a schedule (called 'repertoire') has to be created, which defines the succession of the gratings and how long they are displayed. These repertoires can be saved as a 'repz' file. It is a simple zip file which includes the gratings (in png-format), their schedule and some internal files which define exposure time and trigger behavior. If many gratings or complicated schedules are required, it might be useful to produce this repertoire by hand instead of using the slow MetroCon2.0.

**D. How to rebuild the Fast-SIM**

We believe that our Fast-SIM setup is an easy-to-reproduce solution to whoever wants to build a high-resolution microscope for imaging of biological samples. This is why we indicate here a short manual and made our files for SLM and Arduino online available. However, the alignment of the setup is not completely trivial. Thus, we recommend a little bit of experience.

How to connect the devices (see also Appendix E):
1. Connect the SLM via USB to a PC
2. Connect wires to J3 port of the SLM (20 way, 2rows, Molex, 2.54mm Pitch)
    a. Solder load resistor on the SLM input signal and the ADUM (quad-channel digital isolator – ADUM 1402 (Analog Devices, Norwood, USA)) on the SLM output signal
    b. Connect the cables to the specified Arduino pins
3. Connect the Camera to the PC and the trigger in and output via an BNC cable to the Arduino
4. Connect the AOTF control device to the power supply and the two signal cables to the Arduino
5. Connect the AOTF itself and its control unit to the control box

Settings/Programs:
1. Arduino

    Load the attached program to the Arduino ("ArduinoCode"). The software for uploading is available at www.ardunio.cc.
2. SLM

    Staring the SLM with MetroCon 2.0 and upload our repertoire ("[25 10 7;7 -9 25]_phases_3_rep_3.repz")
3. Camera:
    a. We used "HC image live" from the company to use the camera
    b. Chose external "Level trigger mode" → *"Global exposure level trigger mode"*
    c. Set the polarity of in- and output signals to "positive"

Mask:
Fix a stable cardboard on a holder and put it in the pupil plane of the SLM. Mark with a pen the position of the first orders. Take the cardboard out and prick it with a needle at the marked positions. Use a sharp knife to remove the burr from the backside of the cardboard.

Hints:
A logic analyzer (Saleae, San Francisco, CA, USA) is helpful to understand, debug and repair the electronic circuit. Camera and the laser switch can be replaced by other models. However, it is important to remember than the Arduino operate with 5V logic. Changing the repertoires might require changing the Arduino code and vice versa.

# E. Electronic Circuit

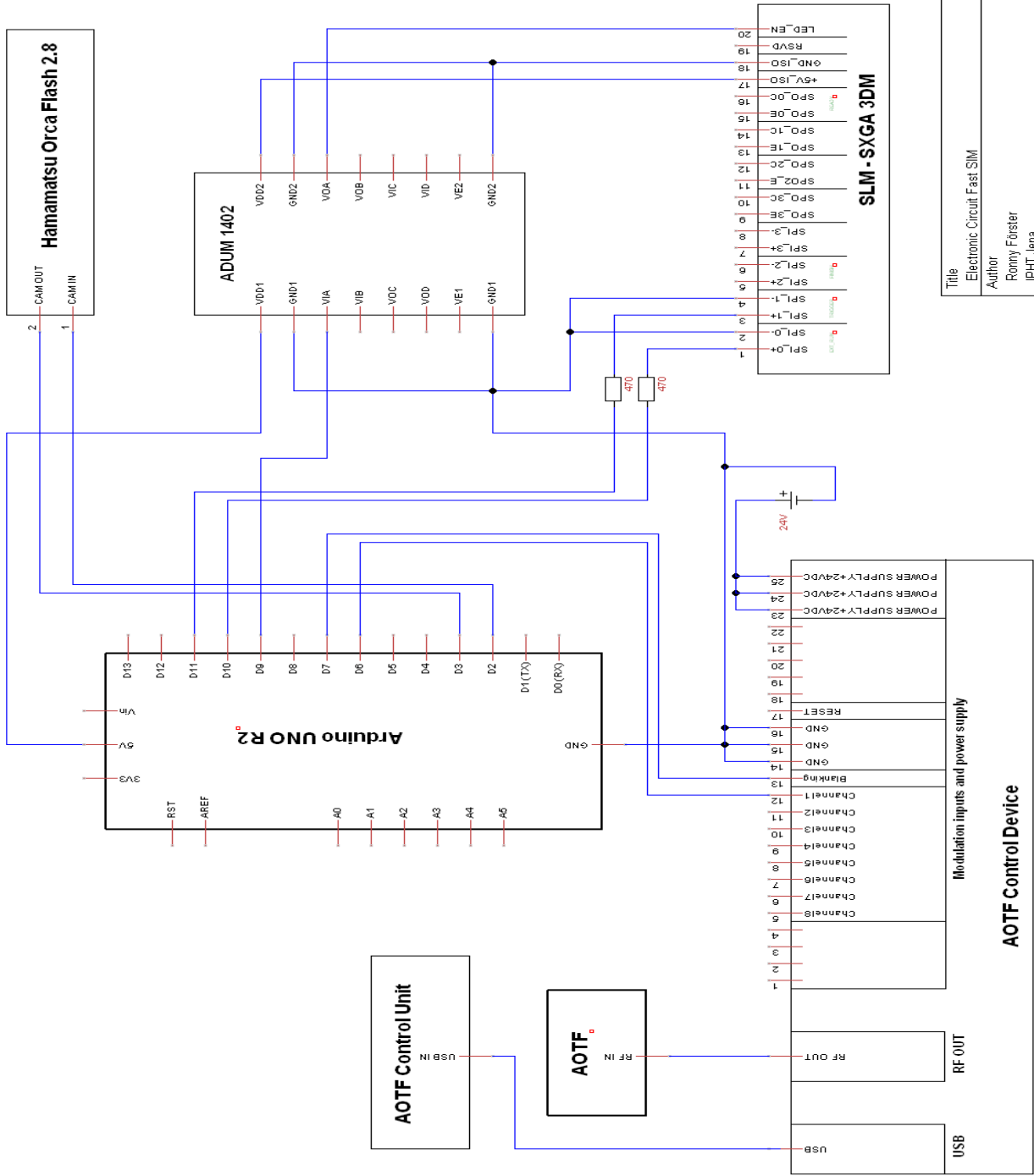

## F. Overview over the online available files (https://github.com/nanoimaging/Appendix-F)

| File/Folder name | info |
|---|---|
| ArduinoCode | Arduino code for synchronization of Fast-SIM system |
| [25 10 7;7 -9 25]_phases_3_rep_3 | Repz file for the SLM with our gratings Load it in MetroCon and send it to the SLM |
| repertoire | Used gratings, sequences and schedule for SIM |
| electronic_circuit | Used electronic circuit for connecting the devices |
| gratings NL SIM | New gratings for NL SIM |

**Acknowledgements**

We thank Robert Kretschmer for the technical support.

**References and links**